\def \Bbar {{\overline {\bm B}}}
\def \Jbar {{\overline {\bm J}}}

\def \Beq {B_{\rm eq}}

\def \A {{\bm A}}

\def \fxt {{\bm f}({\bm x},t)}
\def \f {{\bm f}}

\def \Ubar {{\overline {\bm U}}}

\def \u {{\bm U}}

\def \B {{\bm B}}

\def \j {{\bm J}}

\def \grad {{\bm \nabla}}
\def \curl {{\bm \nabla} \times}
\def \dive {{\bm \nabla}\cdot}
\def \lap {\nabla^2}
\def \delt {\partial_t}
\def \Dt {D_t}

\newcommand{\bra}[1]{\langle #1\rangle}
{}

\def \Rm  {\mbox{Re}_{\rm M}}
\def \Rey  {\mbox{Re}}
\def \omc  {\omega_{\rm c}}

\def \kf  {k_{\rm f}}
\def \kone  {k_{\rm 1}}
\def \urms  {u_{\rm rms}}
\def \lr  {L_{\rm r}}
\def \lt  {L_{\theta}}
\def \lp  {L_{\phi}}

\def \B {\bm B}

\def \A {\bm A}

\def \curl {{\bm \nabla}\times}

\def \etat {\eta_{\rm t}}

\def \Rm  {\mbox{Re}_{\rm M}}

\def \kf  {k_{\rm f}}

\newif\iffigs
\figstrue
\iffigs
\fi
\def\drawing #1 #2 #3 {
\begin{center}
\setlength{\unitlength}{1mm}
\begin{picture}(#1,#2)(0,0)
\put(0,0){\framebox(#1,#2){#3}}
\end{picture}
\end{center} }
\documentclass[preprint2]{emulateapj}
\usepackage{amsfonts,amssymb,amsmath}
\usepackage{graphicx}
\usepackage{bm}
\begin{document}
\title{Oscillatory migrating magnetic fields in helical turbulence in spherical domains}
\author{Dhrubaditya Mitra\altaffilmark{1,2}, Reza Tavakol\altaffilmark{1},
Petri J. K\"apyl\"a\altaffilmark{2,3}, Axel Brandenburg\altaffilmark{2,4}}

\altaffiltext{1}{Astronomy Unit, School of Mathematical Sciences, Queen Mary 
University of London, Mile End Road, London E1 4NS, UK
\email{dhruba.mitra@gmail.com}
}
\altaffiltext{2}{NORDITA, AlbaNova University Center, Roslagstullsbacken 23,
SE-10691 Stockholm, Sweden}
\altaffiltext{3}{Department of Physics, Gustaf H\"allstr\"omin katu 2a 
(PO Box 64), FI-00014, University of Helsinki, Finland}

\altaffiltext{4}{Department of Astronomy, AlbaNova University Center,
Stockholm University, SE-10691 Stockholm, Sweden}
\date{}

\begin{abstract}
We present direct numerical simulations 
of the equations of compressible magnetohydrodynamics
in a wedge-shaped spherical shell, without shear,  
but with random helical forcing which has 
negative (positive) helicity in the 
northern (southern) hemisphere.
We find a large-scale magnetic field that is nearly uniform in the azimuthal 
direction and approximately antisymmetric about the equator.
Furthermore, the large-scale field in each hemisphere oscillates on nearly
dynamical time scales with reversals of polarity and equatorward migration.
Corresponding mean-field models also show similar migratory oscillations
with a frequency that is nearly independent of the magnetic
Reynolds number.
This mechanism may be relevant for understanding
equatorward migration seen in the solar dynamo.

\end{abstract}
\maketitle
%%%%%%%Title page end %%%%%%%%%%%%%%%%%
%%%%%%%%%%% Introduction %%%%%%%%%%%%%%%%%
\section{Introduction}
Large-scale magnetic fields with fascinating 
quasi-regular spatio-temporal behavior 
are ubiquitous in solar and  stellar settings.  
Understanding the mechanisms for the generation of such fields and 
their spatio-temporal variations is still a major challenge for 
dynamo theory. 
The solar magnetic field has three particularly
important features: quasi-regular oscillations, reversal of polarity
and equatorward migration. 
Direct numerical simulations of solar-like convective dynamos
have been able to generate large-scale magnetic fields 
\citep{bro+bro+bru+mie+nel+tom07,brown10}
which in some cases show oscillatory behavior,
but the fields exhibit either rather weak equatorward migration at high
latitudes~\citep{charb} or 
anti-solar (i.e.\ poleward) migration~\citep{gil83,kapyla_etal2010}. 

A useful tool for studying these dynamical phenomena is mean-field (MF)
electrodynamics \citep[e.g.][]{Kra+Rad80,bra+sub05}, where the effects
of turbulence are characterized by turbulent magnetic diffusion
and an $\alpha$ effect.
According to MF theory, equatorward migration
is expected if there is negative radial shear accompanied by a
positive (negative) $\alpha$ effect in the northern (southern) 
hemisphere \citep{Kra+Rad80}.
Direct numerical simulations (DNS) of helical turbulence with shear 
have confirmed the presence of migratory dynamo waves 
\citep{bra+big+sub02,kap+bra09}. It is, however, 
unclear whether this is really what is going on in the Sun,
since there the layer with negative radial shear is rather thin and
only concentrated near the surface (see, e.g., \cite{bra05} and references therein).
The other alternative is that meridional circulation might change the
direction of migration \citep{cho+sch+dik95}, but evidence for
this has not yet been seen in DNS.

In this Letter we present a completely different mechanism for
polarity reversal and equatorward migration of dynamo activity.
In the context of MF models this mechanism is connected with the 
antisymmetry of the profiles
of $\alpha$ across the equator \citep{Rud+Hol04,bra+can+cha09}.
We demonstrate the operation
of this mechanism in DNS of the equations of magnetohydrodynamics (MHD).
Our model consists of a spherical wedge-shaped
shell in which the turbulence in the fluid is maintained by a random
helical forcing.
Motivated by the Sun, we choose our forcing to have opposite
signs of helicity in the two hemispheres
(negative in the north and positive in the south).
We emphasize that, even though our model
does not explicitly include convection, stratification and 
rotation,
the helical forcing used here does partially
model these features implicitly.

\begin{figure}[h]
\includegraphics[width=\linewidth]{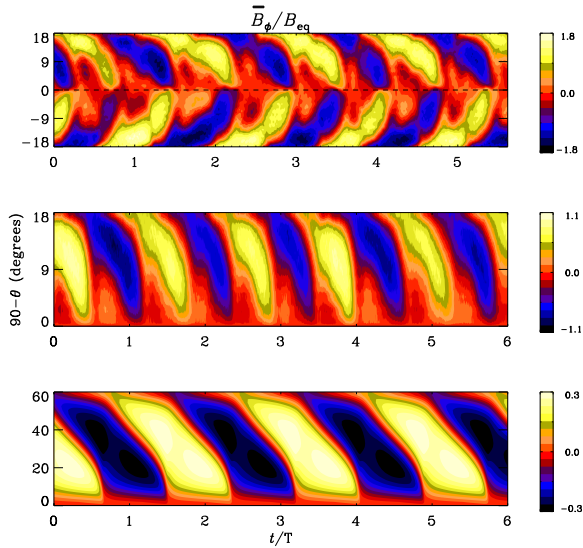}
\caption{Space-time diagrams of the azimuthal component
of the large-scale magnetic field for
DNS over both the hemispheres (top panel),
DNS over only the northern hemisphere with an antisymmetric
condition at the equator (middle panel), and
MF simulation over only the northern hemisphere with 
an antisymmetry condition at the equator (bottom panel).}
\label{fig:butterfly}
\end{figure}

Our model shows large-scale magnetic fields
in excess of the equipartition value. More importantly,
we find oscillations of the magnetic field which show opposite
signs in different hemispheres with periodic reversals of polarity.
Furthermore, the magnetic
field develops at higher latitudes and migrates equatorward where the two
different polarities of magnetic field annihilate and the cycle repeats itself
as shown in the top panel of Fig.~\ref{fig:butterfly}. 
To our knowledge, such
dynamical features of the large-scale magnetic field have not been observed
earlier in DNS of MHD turbulence.
Below we introduce our model, discuss its oscillatory solutions,
and briefly compare our DNS results with those obtained
from corresponding MF models.

%%%%%%%%%%%%%%%%%%%%%%%%%%%%%%%%%%%%%%%%
%-----------------------
\begin{figure}[h]
\includegraphics[width=\linewidth]{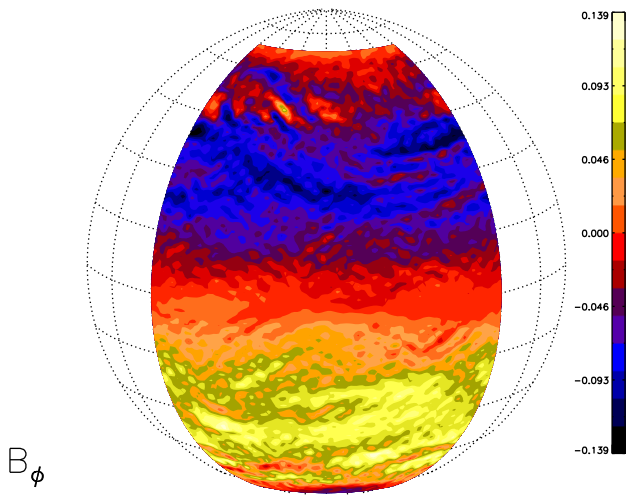}
\caption{Orthographic projection of the toroidal magnetic field $B_\phi$
at $r=0.85$ for Run~{\tt S5}.
The projection is tilted by $15$ degrees towards the viewer. 
} 
\label{fig:contour}
\end{figure}
%%%%%%%%%%%%%%%% Model %%%%%%%%%%%%%%%%%%%%%
\section{The model}
In our simulations, we solve the equations for compressible MHD
in terms of the velocity $\u$, the logarithmic density $\ln\rho$,
and the magnetic vector potential $\A$,
\begin{eqnarray}
\label{mhd1}
\Dt\u &=& -c_{\rm s}^2\grad\ln\rho + \frac{1}{\rho}\j\times\B 
 + \bm{F}_{\rm visc} + \f, \\
\Dt\ln\rho &=& -\grad\cdot\u, \\
\delt\A &= & \u\times\B + \eta\lap\A ,
\end{eqnarray}
where $\bm{F}_{\rm visc}=(\mu/\rho)(\lap\u + \frac{1}{3}\grad\dive\u)$
is the viscous force, $\mu$ is the dynamic viscosity, 
$\B = \curl \A$ is the magnetic field,
$\j = \curl \B/\mu_0$ is the current density,
$\mu_0$ is the vacuum permeability,
$c_{\rm s}$ is the (constant) speed of sound in the medium,
$\eta$ is the magnetic diffusivity,
and $D_t \equiv \delt + \u\cdot\grad$ is the advective derivative.
Our computational domain is a spherical wedge with 
radius $r \in [r_1,r_2]$ symmetric about the equator with 
colatitude $\theta \in [\Theta,\pi-\Theta]$
and azimuth $\phi \in [0,\Phi]$. 
The radial, meridional and azimuthal extents of our domain are respectively,
$\lr \equiv r_2-r_1$, $\lt \equiv r_2(\pi-2\Theta)$ and
$\lp \equiv r_2 \Phi$. All lengths are measured
in the units of $r_2$.
Our main results do not depend on our choices of $\Theta$ and $\Phi$.

In Equation~\ref{mhd1} $\fxt$ is an external white-in-time random helical forcing 
constructed using
the Chandrasekhar-Kendall functions \citep{cha+ken57} as described below.
In spherical coordinates a helical vector function can be expressed
in terms of a scalar potential function:
\begin{equation}
\psi\big(\beta(t),\ell(t),m(t)\big)=
      \mbox{Re}\, z_\ell(\beta r)Y^m_\ell(\theta,\phi)\exp[{\rm i}\xi_m(t)],
\label{solhelmoholtz}
\end{equation}
with 
$z_\ell(\beta r)=a_\ell j_\ell(\beta r) + b_\ell n_\ell(\beta r) $.
Here $j_\ell$ and $n_\ell$ are spherical Bessel functions of the first and
second kind respectively, $a_\ell$ and $b_\ell$ are constants determined by the 
boundary conditions and
$\xi_m$ is a random angle uniformly distributed between $0$ and $2\pi$.
The helical forcing $\f$ is then given by the equation $\curl \f = \beta \f $,
where $\f = {\bm T} + {\bm S}$,
${\bm T} = \curl({\bm e}\psi)$,
and ${\bm S} = \beta^{-1}\curl{\bm T}$,
where ${\bm e}(t)$ is a unit vector chosen randomly on the unit sphere. 
As to the choice of boundary conditions, we demand that $\f$ is zero at the 
two radial boundaries $r=r_1$ and $r=r_2$ which yields the following
transcendental equation relating $a_\ell$, $b_\ell$ and $\beta$:
\begin{equation}
a_\ell j_\ell(\beta r_1) + b_\ell n_\ell(\beta r_1)
= a_\ell j_\ell(\beta r_2) + b_\ell n_\ell(\beta r_2) = 0.
\label{alphaeqn}
\end{equation}
We first construct a table of values of $m$, $\ell$ and $\beta$ in the
following way. 
As we use periodic boundary conditions along the azimuthal direction,
$m_{\rm min}= 2\pi/\Phi$. We choose $m = p m_{\rm min}$, and for a fixed
$m$ we choose $\ell$ to be odd, $\ell = 2(m+q)+1$, because we want the forcing
to go to zero at the equator. 
Here $p$ and $q$ are integers which range between $3$ to $5$.
For a fixed $\ell$ and $m$ we solve Eq.~(\ref{alphaeqn}) by Newton-Raphson
method and list the solutions which have $3$ to $5$ zeros in the domain. 
To randomize the resulting forcing we randomly choose a triplet 
of $m$, $\ell$ and $\beta$
from the table. We also randomize the unit vector ${\bm e}$ on the unit sphere.
Two different signs of helicity are imposed by choosing negative 
(positive) $\beta$
in the northern (southern) hemisphere. The choice of parameters
implies that we have a scale separation between $3$ to $5$ in our simulations.
Our results are fairly robust under the change of different parameters
of forcing. We need scale separation of 3 or more to excite a 
large-scale dynamo \citep{HBD04},
which invariably shows oscillations and equatorward migration. 
Our simulations are performed using the 
\textsc{Pencil Code}\footnote{{\tt http://pencil-code.googlecode.com}};
see \cite{mit+tav+bra+mos08} for details
regarding the implementation of spherical polar coordinates.
%%%%%%%%%%%%%%%%%%%%%%%%%%%%%%%%
\begin{deluxetable*}{c|c|c|c|c|c|c|c|c|c|c|c|c|c}
%\framebox{\begin{tabular}
%Run & Grid & $\lt$  & $\lp$  & $\Bbar_{\rm rms}/B_{\rm eq}$  & $\Rey$ & $\Rm$ &
% $\kf/\kone$ & $\nu\times10^{-5}$ & $\eta\times10^{-5}$ & $\etat\times10^{-5}$&
%$\omc\times10^{-3}$ &$T\times10^{3}$ & $t_{\rm max}$ \\
Run & Grid & $\lt$  & $\lp$  & $\Bbar_{\rm rms}/B_{\rm eq}$  & $\Rey$ & $\Rm$ &
 $\kf/\kone$ & $\nu\times10^{5}$ & $\eta\times10^{5}$ & $\etat\times10^{5}$&
$\omc\times10^{3}$ &$T\times10^{-3}$ & $t_{\rm max}$ \\
\hline
{\tt S1} &$32\times 64\times 32$ &$\pi/5$ & $\pi/10$  & $0.88$  & $5$ & $12$ &
$3$ & $5$ & $2$ & $5.3$ & $2.5$ & $0.18$ & $\sim10T$ \\
{\tt S2} &$64\times 128\times 64$ &$\pi/5$ &$\pi/10$  & $0.79$  & $8$ & $21$ & 
$4$ & $3$ & $1.2$ & $6.2$ & $2.5$ & $0.16$ &  $\sim5T$\\
{\tt S3} &$32\times 64\times 64$ &$\pi/5$ & $\pi/5$ &  $1.16$ & $2$ & $4$ &
$7$ & $5$ & $2$ & $3.6$ & $2$ & $0.27$ & $\sim10T$\\
{\tt S4} &$64\times 128\times 128$ & $\pi/5$ & $\pi/5$ & $1.04$  & $4$ & $10$ & 
$7$ & $2$ & $1$ & $4.9$ & $2.6$ & $0.2$ & $\sim5T$\\
{\tt S5} &$64\times 128\times 128$ & $9\pi/10$ & $\pi/2$ & $2.03$  & $7$ & $13$ & 
$7$ & $2$ & $1$ & $4.4$ & $-$ & $-$ & $\sim T$. \\
%\end{tabular}}
\caption{Summary of our parameters including 
grid size, the meridional and azimuthal extents of our domain, 
rms value of the azimuthally averaged field,
Reynolds number $\Rey$ and  
magnetic Reynolds number $\Rm$.
The forcing amplitude, $f_{\rm amp}=0.2$, is chosen such that the Mach number is 
of the order of 0.1, making the flow essentially incompressible. $t_{\rm max}$
the duration of each run. The run {\tt S5}
has not been ran long enough to accurately measure $\omc$.} 
\label{paratable}
\end{deluxetable*}
%%%%%%%%%%%%%%%%%%

We use periodic boundary conditions along the azimuthal direction
and set the normal component of the magnetic field to zero on all
other boundaries (perfect-conductor boundary condition). 
This is implemented by setting the two tangential components of $\A$
to zero.
As an estimate of the characteristic Fourier mode 
of the forcing we define $\kf=w_{\rm rms}/\urms$,
(column 8 of Table~\ref{paratable})
where $w_{\rm rms}$ and $\urms$ are the rms values of small-scale
vorticity and velocity, respectively.
We introduce the fluid and magnetic Reynolds numbers 
as $\Rey=\urms/\nu\kf$ and $\Rm=\urms/\eta\kf$, respectively. 
Here, $\nu = \mu/\rho_0$ is the mean kinematic
viscosity, where $\rho_0$ is the
initial and the mean density in the volume.
A representative list of parameters is given in Table~\ref{paratable}.
%%%%%%%%%%%%%%%% Results %%%%%%%%%%%%%%%%%
\section{Results}
\begin{figure}
\includegraphics[width=\linewidth]{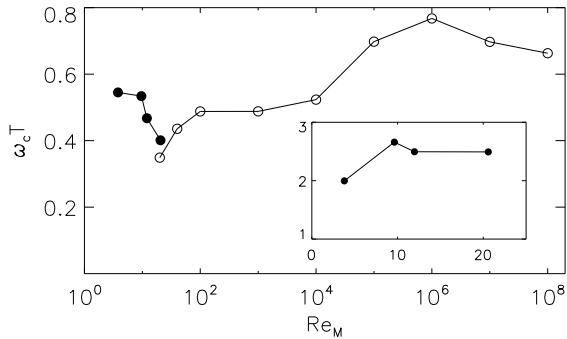}
\caption{
The frequency of the oscillations multiplied
by the turbulent diffusion time $T=(\eta_tk^2_{\theta})^{-1}$ 
as a function of $\Rm$ 
for $\lt = \pi/5 $.
Data from DNS (closed circles) and MF models (open circles) are shown.
For the MF runs $\etat=1$, 
for the DNS runs $\etat$ is given in 
Table~\ref{paratable}. 
The inset shows $\omc \times10^{3}$ (column 12 Table~\ref{paratable})
as a function of $\Rm$ for the DNS data.    
}\label{fig:osc+helm}
\end{figure}

We start our simulations with a random seed magnetic field
of no particular parity about the equator.
After a transient time, of about one turbulent diffusion time, 
we find
that a large-scale magnetic field is generated with energy of the order 
of or exceeding the equipartition strength in all runs.
The  magnetic field encompasses the whole azimuthal extent of 
the domain. 
A contour plot of the toroidal component of the magnetic field 
on a surface with constant radius from Run~{\tt S5} is shown in 
Fig.~\ref{fig:contour}.
We define the large-scale magnetic field via averages over the azimuthal 
and radial directions, i.e.,
$\Bbar = \bra{{\bm B}}_{r\phi}$,
such that the resultant magnetic field is solely a function 
of latitude and time.
We normalize the magnetic field with the equipartition field strength,
$B_{\rm eq}=\bra{\mu_0\rho {\bm u}^2}^{1/2}$,
where ${\bm u} = \u-\Ubar$ is the small-scale velocity. 
The field first develops at higher 
latitudes and then with time migrates equatorward.
In each hemisphere the field shows oscillations
and reversals of polarity. These features can be seen in the space-time diagram
shown in the top panel of Fig.~\ref{fig:butterfly} where we plot
$\overline{B}_\phi$ as a function of latitude and time.
The principal frequency 
of oscillations, $\omc$ (column 12 of Table~\ref{paratable} and the
inset of Fig.~\ref{fig:osc+helm})
is obtained by Fourier transforming the 
time series of ${\overline B}_{\phi}(\theta,t)$ in time at a 
given $\theta$ ($\theta=\pi/20$ say)
and determining the frequency corresponding to the dominant mode.
Normalized energy in the large-scale magnetic field also shows  
oscillations as a function of time, but with frequency
$ 2 \omc$. 
A characteristic dynamical time scale
is the turbulent diffusion time corresponding to the length scale $\lt$ defined by
$T\equiv (k_{\theta}^2 \etat)^{-1}$ where $k_{\theta} = 2\pi/\lt$ and for $\etat$
we take the expression from the first-order-smoothing approximation,
$\etat=\urms/3\kf$. In all our runs we find the product $\omc T$ to be 
of order unity (see the bottom panel of Fig.~\ref{fig:osc+helm}).
%%%%%%%%%%%%%%%%%%%%%%%%%%%%%%%%%%%%%%%%
%%%%%%%%%%%%%%%%%%%%%%%%%%%%%%%%%%%%%%%%

We note that the radial and azimuthal components of the large-scale
magnetic field are almost antisymmetric about the equator.
This allows a further simplification of our model by simulating only one
half of the domain (e.g., the northern hemisphere), while keeping exactly the
same forcing function (e.g., a forcing that is random, and negatively helical 
in the northern hemisphere going smoothly to zero at the equator), but choosing
the boundary condition $B_r=B_\phi=0$ at the equator. Such simulations 
produce exactly the same oscillations (as can be seen by comparing the top and
the middle panels of Fig.~\ref{fig:butterfly}) as those obtained in the
DNS with both hemispheres. 
This implies that these oscillations can be studied with 
half the number of grid points and appropriately chosen boundary 
conditions at the equator. 

Given the large values of the magnetic Reynolds number in solar/stellar 
settings, an important question is how the frequency $\omc$ scales with 
$\Rm$?
This question cannot be answered from DNS because the  
magnetic Reynolds numbers reached are far from 
the asymptotic limit of large $\Rm$~(see Column 7 
of Table~\ref{paratable})
A way forward is to use analogous MF models.
The appropriate setting would be that of an $\alpha^2$ dynamo
with dynamical $\alpha$ quenching~\citep{BB02}, which incorporates 
conservation of magnetic helicity, given by the equations
\begin{eqnarray}
\partial_t \Bbar &=& \curl (\alpha \Bbar) + (\eta+\etat)\lap \Bbar, \\
\partial_t \alpha &=& -2\eta\kf^2 \left( 
                                    \frac{\alpha \bra{\Bbar^2} -\etat \bra{\Bbar\cdot \Jbar}}{\Beq^2} +
                                    \frac{\alpha-\alpha_{\rm K}}{\Rm/3}
                                   \right) ,
\end{eqnarray}
where $\Rm/3=\etat/\eta$ and $\Beq$ is the equipartition field strength.
We use
$\kf/\kone=6$ in our MF simulations. 
In view of the discussion above, we solve the MF equations in only the northern
hemisphere with appropriate boundary condition at the equator.   
In the MF approach the helical nature of turbulence is modelled by the $\alpha$
coefficient ($\alpha_{\rm K}$).
We choose $\alpha_{\rm K} = g(\theta)\alpha_0$ and $\etat = 1$. 
The profile function $g$ takes positive (negative) values in the northern 
(southern) hemisphere, going smoothly to zero at the equator.
This reflects the fact that, according to MF theory, the kinetic $\alpha$
effect usually has the opposite sign to the mean kinetic helicity.
We have used three different functional forms for $g$, namely
$g = \theta-\pi/2$, $g = \sin(\theta-\pi/2) $ and
$ g = \tanh(\theta-\pi/2)$, without any significant change in our results.
We need $\alpha_0 \geq 16$ to excite a dynamo but once excited the oscillatory
and migratory properties of the dynamo do not depend on $\alpha_0$.  
We use perfect-conductor boundary conditions along the radial direction
and our magnetic Reynolds number (changed by varying $\eta$)
ranges between $ 10 \lesssim \Rm \lesssim 10^{8}$. 
We have also used domains with
larger latitudinal extents than those used in our DNS.

Here we briefly mention a few important outcomes of
our MF results relevant to our discussion above:
(a) Our DNS results -- namely oscillatory behavior as well as
migration towards the equator -- are qualitatively reproduced by the MF
solutions in the range of parameters reported here. 
An example of the space-time
diagram produced by our MF runs is shown in the bottom
panel of Fig.~\ref{fig:butterfly}.
(b) The frequency of oscillations remains almost constant as a 
function of $\Rm$, see Fig.~\ref{fig:osc+helm}. 
The dependence of the oscillation period on $\Rm$
seen in DNS may be related to $\Rm$-dependence of
the turbulent magnetic diffusivity. 
A similar behavior has been observed earlier in Cartesian DNS
\citep{kap+bra09}.
(c) To show the robustness of our results with respect to the
size of the domain, we also studied domain sizes extended
in the meridional direction from $\lt = \pi/5$ to $(178/180)\pi$ 
which corresponds respectively to  $\Theta= 72 $ degrees and $1$ degree.
We find that the oscillations and the migratory behavior do not change.
(d) For the MF model considered here the mean value of 
the large-scale magnetic field decreases as $\Rm^{-1}$, i.e., the field is catastrophically 
quenched for large values of $\Rm$.  
In the DNS, however, such quenching could be alleviated by magnetic 
helicity fluxes across the equator \citep{bra+can+cha09,mitra_etal2010}.

To test the robustness of our simulations with respect 
to the choice of boundary condition in the radial direction 
we repeated our simulations with the vertical field boundary 
condition, which makes the two tangential components of the magnetic 
field vanish at the radially outward boundary. 
These simulations also show oscillations and
equatorward migration of magnetic activity, but in this case the
oscillations are less regular and the frequency is marginally higher. 
%%%%%%%%%%%%%%%% Conclusion %%%%%%%%%%%%%%%%%
\section{Conclusions}
We have found large-scale fields, oscillations on
dynamical time scales and polarity reversals with equatorward
migration of magnetic activity 
in direct numerical simulations of helically
forced MHD equations in spherical wedge domains. 
Despite its simplicity, it is quite
striking how our model can reproduce these important features of
the solar dynamo. As far as we are aware, these features 
have not been observed earlier in DNS.
We have elucidated our DNS results by considering
analogous $\alpha^2$ MF models which support
our conclusions. 
We have further used these MF models
to explore magnetic Reynolds numbers that are 
at present inaccessible to DNS. 
This has enabled us to show that the frequency 
of the oscillations is almost
independent of $\Rm$
for large $\Rm$.
Such MF models have been known to have oscillatory solutions
if $\alpha$ changes sign in the computational domain 
\citep[see, e.g.][]{bar+shu87,ste+ger03,Rud+Hol04,bra+can+cha09},
but their migratory property had not been studied before.
Antisymmetry of $\alpha$ with depth also produces oscillatory 
solutions \citep{bar+shu87,ste+ger03}, but not the equatorward
migration. 

The helical forcing used in our DNS, with its different
signs of helicity in different hemispheres, implicitly 
models only the helical aspect of the effects of rotation 
and stratification present in the Sun. 
Physically, a more complete picture should emerge from DNS of convective
turbulent dynamo as done, for example, by 
\cite{gil83,bro+bro+bru+mie+nel+tom07,brown10,kapyla_etal2010,charb}.
Such simulations also generate differential rotation and lead
to another dynamo mode of operation -- the $\alpha\Omega$ dynamo
-- which also produces oscillatory behavior.
However, in order to get equatorward
migration radial shear must be negative. 
Helioseismology has shown that negative radial shear exists only
near the surface of the convection zone.
This feature has so far not been reproduced by global DNS.
Whether or not an $\alpha\Omega$ dynamo is the dominant mechanism operating
in the Sun remains unclear.
It is therefore important to keep in mind 
that there exists alternative mechanisms 
for producing oscillatory behavior with 
equatorward migration, such as the one discussed here.
%---------------------
\acknowledgments
%--------------------
This work was supported in part by
the European Research Council under the AstroDyn Research Project No.\ 227952
and the Swedish Research Council Grant No.\ 621-2007-4064.
DM is supported by the Leverhulme Trust.
PJK is supported by the Academy of Finland grant No.\ 121431.
Computational resources were granted by
UKMHD, QMUL HPC facilities purchased under the SRIF initiative,
the National Supercomputer Centre in Link\"oping in Sweden, and 
CSC--IT Center for Science in Espoo, Finland.

%%%
\newcommand{\ypre}[3]{ #1, {Phys.\ Rev.\ E,} {#2}, #3}
%%%%%%%%%%%%%%%%%%%%%%%%%%%%%%%%%%%%% 

%%%%%%%%%%%%%%%%%%%%%%%%%%%%%%%%%%%%%%%%%
\end{document}